\documentclass[%
 reprint,
superscriptaddress,
 amsmath,amssymb,
 aps,
floatfix,
]{revtex4-2}

\usepackage{graphicx}
\usepackage{dcolumn}
\usepackage{bm}
\usepackage{amsmath,amssymb}
\usepackage{color}
\usepackage{textcomp,mathcomp}

\begin{document}

\preprint{APS/123-QED}

\title{Metastable ordered states induced by low temperature annealing of $\delta$-Ag$_{2/3}$V$_2$O$_5$ }

\author{T. Kubo}
\affiliation{Department of Applied Physics, Nagoya University, Aichi 464-8603, Japan}
\author{K. Kojima}
\affiliation{Department of Applied Physics, Nagoya University, Aichi 464-8603, Japan}

\author{N. Katayama}\thanks{Corresponding author}\email{katayama.naoyuki.m5@f.mail.nagoya-u.ac.jp}
\affiliation{Department of Applied Physics, Nagoya University, Aichi 464-8603, Japan}

\author{T. Run\v{c}evski}
\affiliation{Department of Chemistry, Southern Methodist University, 225 University Blvd., Dallas TX 75205, U.S.A.}


\author{R. E. Dinnebier}
\affiliation{Max Planck Institute for Solid State Research (MPI-FKF), Heisenbergstraße 1, 70569 Stuttgart, Germany}

\author{A. S. Gibbs}
\affiliation{Max Planck Institute for Solid State Research (MPI-FKF), Heisenbergstraße 1, 70569 Stuttgart, Germany}
\affiliation{ISIS Neutron and Muon Source, Rutherford Appleton Laboratory, Chilton, Didcot, OX11 0QX, United Kingdom}

\author{M. Isobe}
\affiliation{Max Planck Institute for Solid State Research (MPI-FKF), Heisenbergstraße 1, 70569 Stuttgart, Germany}

\author{H. Sawa}					
\affiliation{Department of Applied Physics, Nagoya University, Aichi 464-8603, Japan}
\date{\today}

\begin{abstract}
In $\delta$-Ag$_{2/3}$V$_2$O$_5$ with charge degrees of freedom in V, it is known that the charge ordering state and physical properties of V that appear at low temperatures depend strongly on the ordering state of Ag. In this study, we focused on the Ag ions in the interlayer and studied the structure using synchrotron radiation powder diffraction in dependence on temperature. We found that when the sample is slowly cooled from room temperature and ordering occurs at the Ag sites, V$^{4+}$/V$^{5+}$ charge ordering of V and subsequent V$^{4+}$-V$^{4+}$ structural dimers are produced. Although quenching the sample from room temperature suppresses the ordering of Ag, annealing at around 160 K promotes partial ordering of Ag and allows a metastable phase to be realized. This metastable phase is maintained even when the temperature is lowered again, producing a remarkable change in low-temperature properties. These results indicate that the ordered state of Ag, which is the key to control the charge-ordered state and physical properties, can be controlled by low-temperature annealing. The results of this study may provide a methodology for the realization of metastable states in a wide range of material groups of vanadium compounds, where competition among various charge ordered states underlies the physical properties.

\end{abstract}

\maketitle


\section{\label{sec:level1}Introduction}
The search for various ground states in vanadium compounds with charge degrees of freedom is one of the major research topics in condensed matter physics. Examples include the heavy fermion states in LiV$_2$O$_4$ \cite{LiV2O4, LiV2O4-2, LiV2O4-3, LiV2O4-4, LiV2O4-5}, the appearance of superconducting phases adjacent to charge ordered phases in $\beta$-Na$_{0.33}$V$_2$O$_5$ \cite{Na0.33V2O5, Na0.33V2O5-2, Na0.33V2O5-3}, various charge ordered states called ``devil's flower'' appearing in NaV$_2$O$_5$ under pressure \cite{NaV2O5, NaV2O5-2, NaV2O5-3}, various molecular formation patterns in the Li$_x$V$X$$_2$ system (0 $\leqq$ $x$ $\leqq$ 1, $X$ = O, S, Se) \cite{Goodenough, Tian, LiVO2-kj, LiVO2-kj2, LiVSe2, LiVS2-1, LiVS2-2, Li033VS2, LiVS2_NMR_Tanaka1, LiVS2_NMR_Tanaka2, LiVO2_NMR_Kawasaki}, and the recently discovered unconventional superconductivity emerging from charge ordered states in CsV$_3$Sb$_5$ \cite{CsV3Sb5, CsV3Sb5-2, CsV3Sb5-3, CsV3Sb5-4, CsV3Sb5-5}. Competition between multiple electronic phases due to charge degrees of freedom underlies the variety of ground states realized in these systems. If we can freely extract the competing ground states that universally exist in vanadium compounds with charge degrees of freedom as metastable phases, we will be able to transform existing material systems into a stage for further exploration of physical properties.

$\delta$-Ag$_{2/3}$V$_2$O$_5$ may provide a new playground to freely extract such competing metastable phases \cite{Ag2/3V2O5, Ag2/3V2O5-2}. $\delta$-Ag$_{2/3}$V$_2$O$_5$ has a layered structure with a non-integer valence of V$^{4.66+}$. Although Ag is apparently present at one site, single-crystal X-ray diffraction results report that it is actually split into two sites with slight local atomic displacement \cite{Onoda}. As the temperature gradually decreases, below about 225 K, the Ag$^+$ sites become three sites, with significant shading in the occupancy of each Ag site. This ordering of the Ag sites is believed to cause gradation in the Coulomb potential on the vanadium sites, separating the valence of V$^{4.66+}$ into V$^{4+}$ and V$^{5+}$. A further decrease in temperature leads to a sharp decrease in magnetic susceptibility at temperatures below 100 K. At low temperatures, spin gap states are expected to be realized \cite{Ag2/3V2O5, Ag2/3V2O5-2}. However, structural changes suggestive of dimer formation have not been captured so far.
 
Interestingly, quenching from high temperatures suppresses the ordering of Ag, allowing it to freeze while maintaining the arrangement of Ag at high temperatures. It has been reported that such quenched samples do not exhibit a decrease in magnetic susceptibility below 100 K like slow-cooled samples and remain paramagnetic down to the lowest temperature. When the temperature of the quenched sample is slowly increased, the Ag frozen at the site begins to melt and forms an ordered state. Further increasing the temperature causes a return to the disordered state. Then, is it possible to control the melting process of frozen Ag to create various metastable ordered states? The realization of a metastable Ag arrangement should induce a change in the charge-ordered state of V, which is strongly correlated with the Ag$^{+}$ arrangement, and produces various low-temperature properties that should not appear originally.

In this paper, we report the following three points obtained from our synchrotron X-ray structural studies of $\delta$-Ag$_{2/3}$V$_2$O$_5$. First, the local displacement of Ag sites present at high temperatures disappears as soon as slow cooling leads to ordering of Ag site occupancy. We show that the appearance and disappearance of local displacements can be explained with the effect of Coulomb repulsion between Ag ions. Second, ordering of Ag leads to charge ordering of V$^{4+}$/V$^{5+}$, followed by structural dimerization associated with spin singlet between the adjacent V$^{4+}$ ions below 100 K, which were observed from our synchrotron X-ray structural analysis. Third, annealing of the quenched sample at around 160 K induces a partial ordering of Ag that depends on the annealing temperature, and the Ag state thus obtained is maintained at low temperatures. These metastable phases yield different physical properties than either the slow-cooled or quenched samples. This result reveals the ground state of $\delta$-Ag$_{2/3}$V$_2$O$_5$ from the structural aspect from a comprehensive structural study using quantum beams, and proposes a methodology for extracting metastable phases from the competing charge-ordered states of vanadates. The latter can be a useful and universal idea for various material systems.

\section{\label{sec:level2}Results and discussions}
\subsection{\label{sec:levelA}Sample Preparation and experimental details}
$\delta$-Ag$_{2/3}$V$_2$O$_5$ powder samples were obtained by mixing Ag$_2$O, V$_2$O$_3$ and V$_2$O$_5$ in stoichiometric ratio, pelletizing, vacuum-sealing in quartz tubes, and sintering at 500 °C for 1 day. Magnetization measurements were performed using the Quantum Design MPMS at the Institute of Solid State Physics, University of Tokyo, in the temperature range from 4 K to 300 K in three steps (2 K/min, 5 K/min, 10 K/min) under a 1 T magnetic field. Quenching was performed by quickly inserting a room temperature sample into the MPMS sample chamber, which was cooled to approximately 5 K. Powder X-ray diffraction experiments were performed at BL02B2 and BL19B2 at SPring-8, BL5S2 at Aichi SR and X17B1 at the National Synchrotron Light Source (NSLS) of the Brookhaven National Laboratory (BNL). At BL02B2 at SPring-8, diffraction experiments were performed in the temperature range from 30 K to 300 K using 23 keV X-rays and changing the temperature at a rate of 10 K/min with a He gas blowing device. At BL02B2 at SPring-8, X-ray diffraction experiments were performed on well-milled fine powders using 25 keV X-rays. At BL5S2 of Aichi SR, samples were quenched by inserting a glass capillary in N$_2$ gas set at 100 K. X-ray diffraction measurements were performed using 20 keV X-rays. At X17B1 of NSLS, X-ray diffraction experiments were performed using 67.42 keV X-rays. Diffracted X-rays were detected using a Perkin Elmer image plate detector. The measured sample was sealed in a glass capillary of 0.5 mm diameter and cooled/heated with an Oxford cryostream 700 cold gas blower. The integrated measurement time for each frame was 60 seconds. High-resolution neutron diffraction data were collected on the time-of-flight powder diffractometer HRPD at the ISIS Neutron and Muon Source. The data were collected at 300 K and 100 K in a standard time-of-flight window of 30–130 ms with the sample contained in standard vanadium can within a closed-cycle refrigerator. RIETAN-FP \cite{RIETAN} was used for analysis, and VESTA \cite{VESTA} was used to draw crystal structures.

\begin{figure}
\includegraphics[width=86mm]{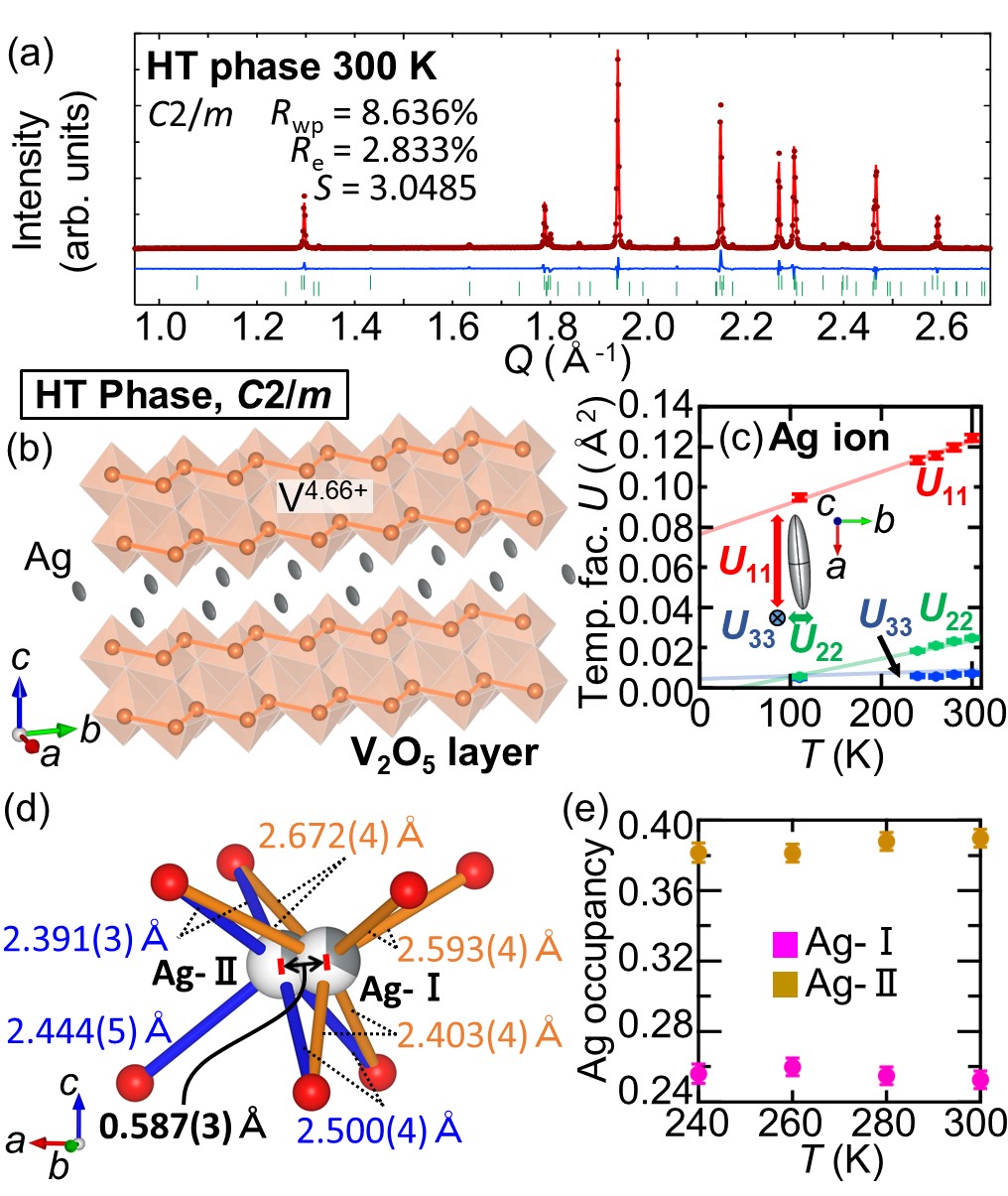}
\caption{\label{fig:fig1} (a) Results of Rietveld analysis of synchrotron X-ray diffraction data in high temperature phase 300 K. Data were collected at the BL02B2 beamline at SPring-8, and the X-ray energy was 23 keV. (b) Crystal structure of high temperature phase. (c) Temperature dependence of anisotropic thermal parameters $U_{11}$, $U_{22}$, and $U_{33}$. The inset shows the thermally oscillating ellipsoid of an Ag atom. (d) Coordination relationship between Ag and oxygen obtained by Rietveld structural analysis in the Ag two-site model. (e) Temperature dependence of the occupancy at the two Ag sites determined in (d), 6-coordinated Ag-I and 5-coordinated Ag-I\hspace{-1.2pt}I.}
\end{figure}


\subsection{\label{sec:levelB}Confirmation of local atomic displacement of Ag appearing at high temperatures}

Figure~\ref{fig:fig1}(a) shows a result of Rietveld analysis of the diffraction pattern obtained at high temperature of 300 K. Hereafter, the three electronic phases produced by slow cooling are referred to as the ``high temperature(HT) phase'', ``intermediate temperature(IT) phase'', and ``low temperature(LT) phase'', respectively. The data were successfully fitted by assuming $\delta$-Ag$_{2/3}$V$_2$O$_5$ with $C$2/$m$ space group as the main phase and $\beta$-Ag$_{1/3}$V$_2$O$_5$ of about 1.5 \% in molar ratio as an impurity. The obtained main phase structure of $\delta$-Ag$_{2/3}$V$_2$O$_5$ is shown in Figure~\ref{fig:fig1}(b). Assuming that Ag is present at one site, its isotropic temperature factor parameter becomes $B$(\AA$^2$) = 4.34(6), which is unusually large compared to a standard materials such as AgO \cite{AgO}. Therefore, we performed a Rietveld analysis using the anisotropic temperature factor parameter and found that the $a$-axis component, the $U_{11}$ parameter, is anomalously large. The temperature dependence of each parameter of the anisotropic temperature factor for the Ag site is shown in Figure~\ref{fig:fig1}(c). While $U_{22}$ and $U_{33}$ are monotonically decreasing to 0 at 0 K, $U_{11}$ maintains a finite large value even at 0 K. This indicates that the anomaly in the $U_{11}$ parameter is not simply due to thermal oscillation but to the local atomic displacement. 

In the past literature, single-crystal X-ray diffraction experiments have shown that in the high-temperature phase of $\delta$-Ag$_{2/3}$V$_2$O$_5$, Ag is divided into two very close sites \cite{Onoda}. We reanalyzed the data with reference to this previous literature and the diffraction data were successfully fitted by placing two Ag sites with the usual temperature factors at the positions coordinated by 5 and 6 oxygen atoms, as shown in Figure~\ref{fig:fig1}(d). The occupancies are 0.390(5) for the 5-coordinated site and 0.253(5) for the 6-coordinated site, as shown in Figure~\ref{fig:fig1}(e), and the sum of the two occupancies is about 2/3, which is consistent with the ratio of Ag indicated by the composition, so the analysis is considered as reliable. The reason for the difference in occupancy between the 5- and 6-coordination sites was unknown, but we found a valid reason based on the temperature dependence of the local atomic displacement, which is discussed later in the G. Discussion section of this paper.

\subsection{\label{sec:levelC}Ordering of Ag and associated charge ordering of vanadium that appears with slow cooling}

\begin{figure}
\includegraphics[width=86mm]{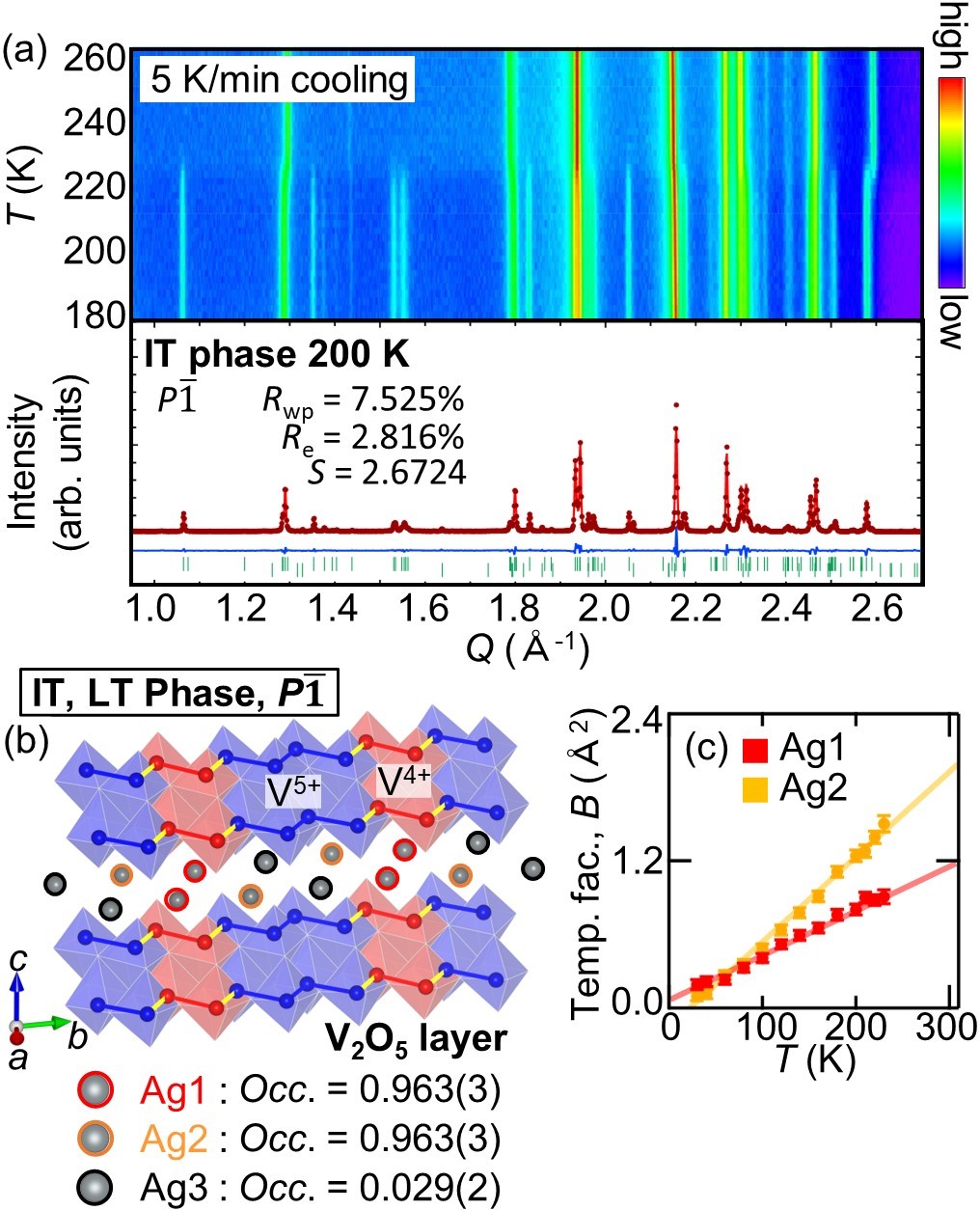}
\caption{\label{fig:fig2} (a) (upper) Color map showing the change in diffraction patterns during the slow cooling process from room temperature. (lower) Results of Rietveld analysis of synchrotron X-ray diffraction data in intermediate temperature phase of 200 K. Data were collected at the BL5S2 beamline at Aichi SR, and the X-ray energy was 20 keV. (b) Crystal structure of intermediate and low temperature phases and occupancy of three Ag sites. (c) Temperature dependences of the isotropic displacement parameters of Ag1 and Ag2 sites. The isotropic displacement parameter of the Ag3 site is not shown in the figure because the occupancy of the Ag3 site is almost zero.}
\end{figure}

When the temperature was slowly lowered, a structural phase transition occurred at about 225 K, and multiple superlattice peaks appeared below the transition temperature. The temperature dependence of the diffraction pattern is summarized in a color map consisting of 80 data obtained by exposing each image for 20 seconds while lowering the temperature at 5 K/min, as shown in Figure~\ref{fig:fig2}(a). Below the transition of 225 K, the Ag sites are split into three sites, Ag1, Ag2, and Ag3, and their respective occupancies were determined to be 0.964(3), 0.964(3), and 0.029(2), as shown in Figure~\ref{fig:fig2}(b). The average occupancy of the three sites was 0.652(3). The total amount of Ag ions in the unit cell in the high and intermediate temperature phases agree within error, indicating that the results of the analysis are reliable. The isotropic displacement parameter for the Ag sites was reduced to a value of $B$ $\textless$ 1.5(\AA$^2$). This $B$ value is the magnitude of the normal displacement parameter, which decreases monotonically with temperature to zero at 0 K, as shown in Figure~\ref{fig:fig2}(c). This indicates that there is no local displacement of Ag sites in the intermediate temperature phase as in the high temperature phase. The results of the precise structure analysis indicate that one of the two Ag ion sites that existed in close proximity in the high temperature phase was selected for ordering.

As mentioned above, when the Ag site splits into three sites with very different occupancies in the intermediate temperature phase, the Coulomb repulsion that each vanadium site receives from the Ag$^+$ ions is also expected to change significantly. To best relieve this Coulomb repulsion from the Ag$^+$ ions, vanadium is argued to undergo charge ordering to V$^{4+}$ (3$d^1$) and V$^{5+}$ (3$d^0$) in the intermediate temperature phase. As a consequence of the low symmetrization associated with the phase transition, the number of vanadium sites increases from two in the high-temperature phase to six in the intermediate temperature phase. Based on the occupancy of Ag sites in the intermediate temperature phase, the six V sites are expected to be divided into two V$^{4+}$ sites shown in red and four V$^{5+}$ sites shown in blue, as shown in Figure~\ref{fig:fig2}(b).

\subsection{\label{sec:levelD}Periodic formation of V$^{4+}$-V$^{4+}$ dimer with spin-singlet state in the low temperature phase}

\begin{figure}
\includegraphics[width=86mm]{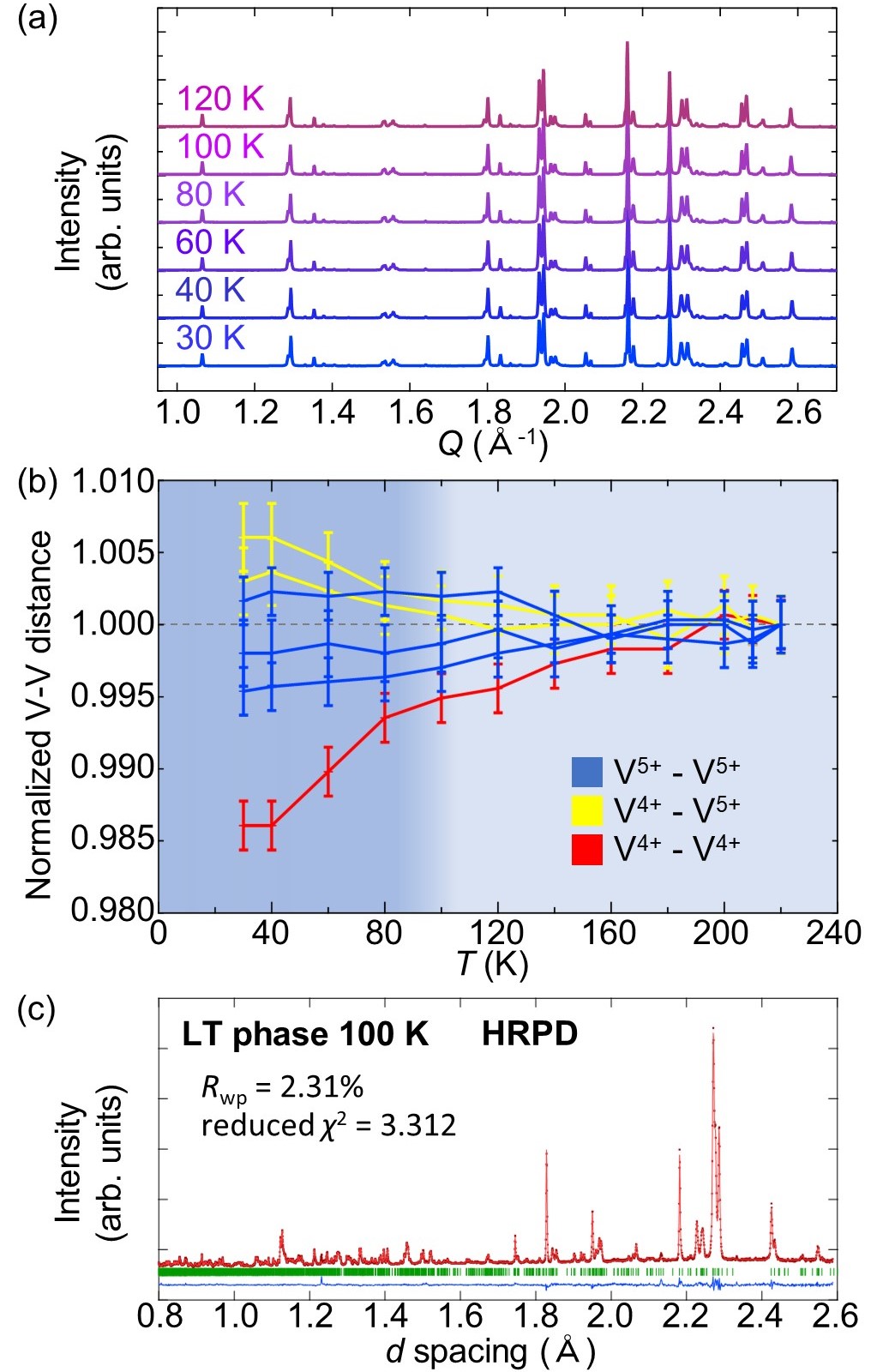}
\caption{\label{fig:fig3} (a) Temperature variation of synchrotron X-ray diffraction patterns from intermediate to low temperature phases. Data were collected at the BL02B2 beamline at SPring-8, and the X-ray energy was 23 keV. (b) Temperature dependence of six different V-V distances normalized at 220 K in the intermediate temperature phase. Blue, yellow and red correspond to the colors of V-V bonds shown in Figure~\ref{fig:fig2}(b), respectively. (c) Results of neutron diffraction experiments performed just below the transition between the low and intermediate temperature phases. }
\end{figure}


As shown in Figure~\ref{fig:fig3}(a), no significant change in the diffraction pattern occurred even when the temperature was lowered below 100 K, where a decrease in magnetic susceptibility appears \cite{Ag2/3V2O5}. Therefore, using the structure of the intermediate temperature phase as an initial structural model, the structure of the low-temperature phase below 100 K could be refined by Rietveld analysis. As shown in Figure~\ref{fig:fig2}(c), the Debye-Waller coefficients for the Ag1 and Ag2 sites monotonically decrease beyond 100 K, the phase transition temperature from the intermediate temperature phase to the low temperature phase, suggesting that the low temperature phase structure has been successfully refined. The obtained structure is similar to the intermediate temperature phase structure at 200 K, but a significant change in the V-V distance is observed below 100 K. In the intermediate and low temperature phases, vanadium becomes six sites, resulting in six nearest-neighbor V-V distances. Figure~\ref{fig:fig3}(b) shows the temperature dependence of the six V-V distances normalized at 220 K in the intermediate temperature phase. The red-blue-yellow color corresponds to the color of the V-V bonding in Figure~\ref{fig:fig2}(b). Only the red V-V distance decreases significantly below 100 K, suggesting the formation of a dimer. Based on the ordered state of Ag, this dimer is expected to be composed of a pair of V$^{4+}$ ions with spin $S$=1/2. Therefore, we can easily expect that a dimer is formed between V$^{4+}$ ions generated by the Ag ordering below 100 K, and that the formation of spin singlet mediated by this dimer is responsible for the decrease in the magnetic susceptibility.

Based on the sharp decrease in magnetic susceptibility, one might consider the possibility of an antiferromagnetic transition originating from the spin $S$ = 1/2 of the V$^{4+}$ ions, but it should be mentioned that this has been ruled out based on $^{51}$V-NMR measurements \cite{Ag2/3V2O5_NMR}. Furthermore, as shown in Figure~\ref{fig:fig3}(c), no magnetic peaks suggestive of antiferromagnetism were observed in the neutron diffraction experiments just below the phase transition.

\subsection{\label{sec:levelE}Quench-induced suppression of Ag site order and continuous change in Ag site occupancy with increasing temperature}

\begin{figure}
\includegraphics[width=86mm]{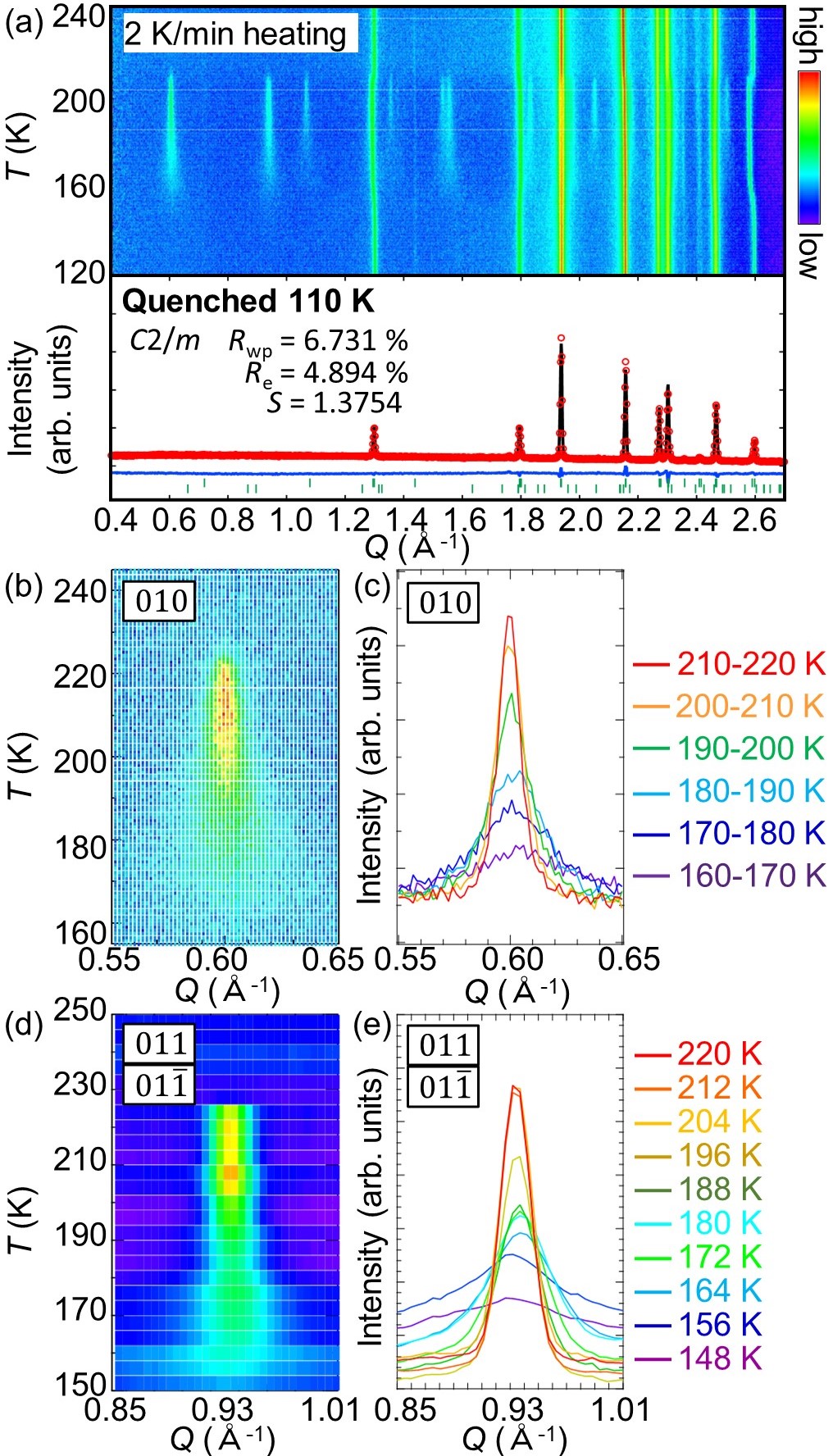}
\caption{\label{fig:fig4} (a) (upper) Color map showing the change in diffraction patterns during the slow heating process of the quenched sample. Data were collected at the BL5S2 beamline at Aichi SR, and the X-ray energy was 20 keV. (lower) Results of Rietveld analysis of synchrotron X-ray diffraction data in quenched sample of 110 K. (b) Enlarged image of 010 superstructure peak. (c) Temperature dependences of 010 peaks. (d) Temperature dependence of the X-ray diffraction pattern using high energy X-rays at 67.42 keV. The development of the 011,01$\bar{1}$ peak intensity is consistent with the 010 peak obtained from measurements at 20 keV, confirming that it is independent of the X-ray energy used. (e) Temperature dependences of 011,01$\bar{1}$ peaks.}
\end{figure}

We have seen how the Ag ordering induced by slow cooling significantly alters the Coulomb repulsion on the vanadium sites, causing charge ordering of vanadium to V$^{4+}$ and V$^{5+}$, followed by dimerization between adjacent V$^{4+}$ ions with spin singlet as the ground state. It has been reported that this Ag ordered state can be suppressed by rapid cooling \cite{Ag2/3V2O5}. In the following, we will show that a metastable Ag ordered state can be induced by gradually melting the frozen Ag state as the temperature rises, thereby inducing a change in low-temperature properties.

The diffraction pattern at 110 K obtained from a sample quenched from room temperature is shown in Figure~\ref{fig:fig4}(a). The N$_2$ blowing gas used to control the temperature for X-ray diffraction was set to 110 K, and quenching was achieved by inserting the sample into the cold gas. The diffraction pattern of the quenched sample at 110 K is similar to that of the high temperature phase obtained at 300 K and can be refined by assuming the same $C$2/$m$ space group as the high temperature phase. Assuming that Ag is located on one crystallographic site and fixing the site occupancy to 0.65, and a Rietveld analysis of the 110 K data with reference to the analysis at 300 K yields a large isotropic Debye-Waller factor $B$(\AA$^2$) = 2.24(8). When anisotropic temperature coefficients were used for the Ag sites, anisotropic elongation was observed in the thermally oscillating ellipsoid, as seen in the inset of Figure~\ref{fig:fig1}(c). Interestingly, as shown in Figure~\ref{fig:fig1}(c), the parameters $U_{11}$, $U_{22}$, and $U_{33}$ at 110 K for the quench sample lie on a straight line extrapolated from these parameters obtained at 240 K to 300 K. Therefore, assuming the existence of two sites, we performed a structural analysis in the same way as we did for the high-temperature data. The occupancies are shown in Figure~\ref{fig:fig1}(e). This is consistent with the fact that the quench completely suppresses Ag ordering and maintains the high-temperature phase structure at 110 K. The temperature dependence of the diffraction pattern of the quenched sample is summarized in a color map consisting of 80 data obtained by exposing each image for 20 seconds while heating the temperature at 2 K/min, as shown in Figure~\ref{fig:fig4}(a). The superlattice peak appears around 160 K and disappears again around 220 K. The diffraction pattern that appears between 160 K and 220 K is basically similar to the diffraction pattern of the intermediate temperature phase that appears below 225 K on slow cooling, and is thought to originate from the ordering of the frozen Ag sites in the quenched sample. As the temperature is further increased, the diffraction pattern changes again, indicating that above 225 K, the ordered state of Ag dissolves and becomes disordered again.

Figures~\ref{fig:fig4}(b) and \ref{fig:fig4}(c) both show the temperature dependence of the 010 superlattice peak that appears above 160 K. Figure~\ref{fig:fig4}(c) was obtained by merging multiple data corresponding to a temperature range of about 10 K, which were continuously measured during the temperature increase. The intensity gradually becomes stronger and the peak width becomes narrower on heating. Note that a similar behavior has been observed in experiments using high-energy X-rays, indicating that the phenomenon is independent of X-ray energy, as shown in Figure~\ref{fig:fig4}(d). This indicates that this structural phase transition has the property of slowly changing as a function of temperature and/or time. The components that produce the intensity of the peak $I_{010}$ can be obtained from the following equations.

\begin{flalign}
&\ \ I_{hkl} \propto |F_{hkl}|^2 &
\end{flalign}
\begin{flalign}
&\ \ F_{hkl} = \sum_j f_j {\rm exp} \{-2 \pi i (hx_j + ky_j + lz_j)\} &
\end{flalign}
\begin{flalign}
\ \ F_{010} = &\ (0.058)f_{\rm V} + (0.147)f_{\rm O} & \notag\\
&+\{(1.921)a_1 - (0.668)a_2 - (1.345)a_3\}f_{\rm Ag} \\
\notag\\
\sim &\ \{(1.921)a_1 - (0.668)a_2 - (1.345)a_3\}f_{\rm Ag}
\end{flalign}

$\newline$
where $F_{hkl}$ is the crystal structure factor of the $hkl$ peak, $f_{\rm V}$, $f_{\rm O}$ and $f_{\rm Ag}$ are the atomic scattering factors of the corresponding elements, and $a_1$, $a_2$, and $a_3$ are the occupancies of the three Ag sites. As shown in Equations (3) and (4), the contributions of oxygen and vanadium to the intensity are very small and the Ag site occupancy is dominant. Therefore, the intensity change of the superlattice peak from 160 K to 220 K suggests a continuous change in Ag site occupancy in this temperature region, and the change in peak width indicates that the correlation length of Ag sites increases with increasing temperature.

\subsection{\label{sec:levelF}Control of Ag ordering process by annealing of quenched samples}

\begin{figure}
\includegraphics[width=86mm]{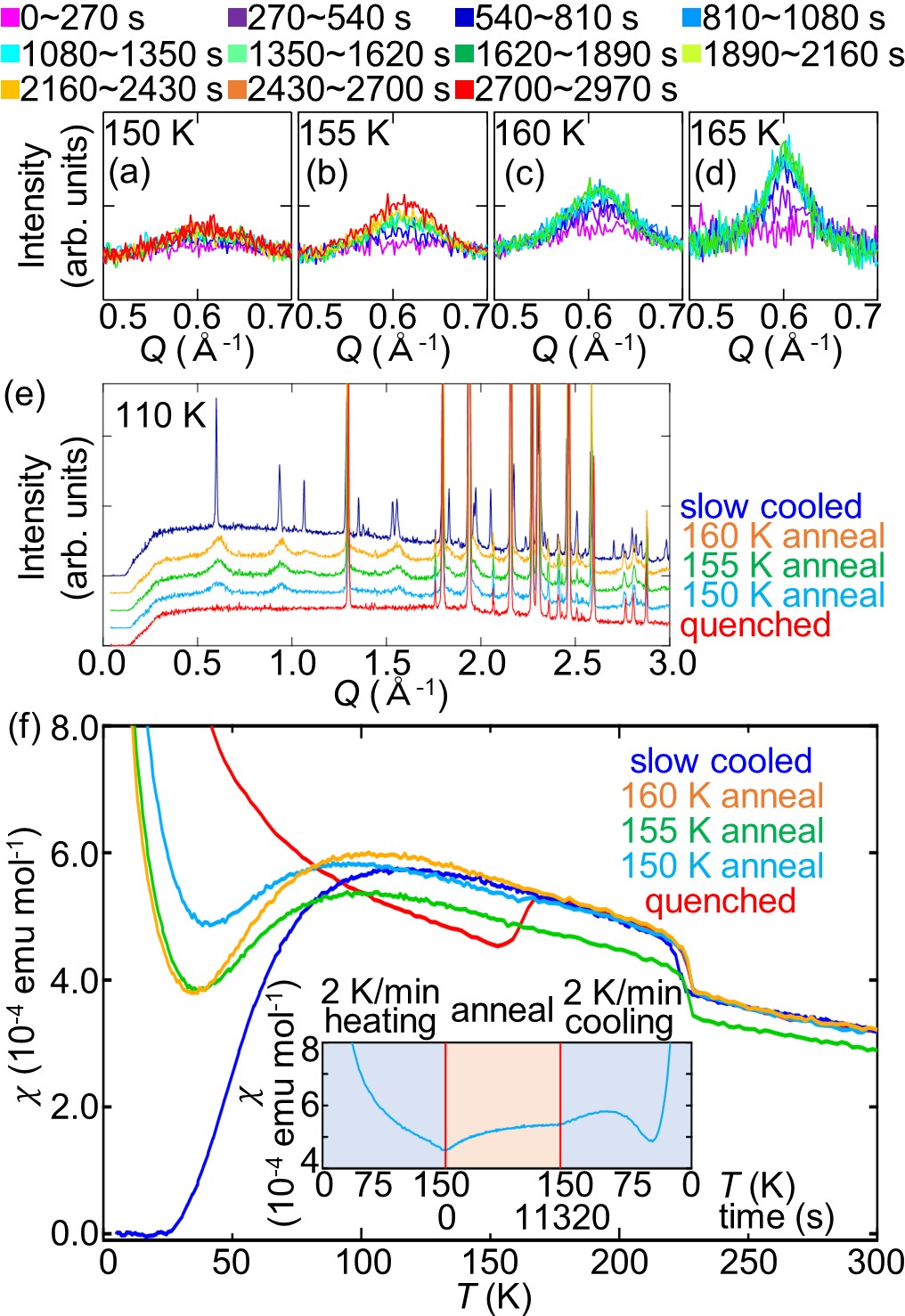}
\caption{\label{fig:fig5} (a-d) Time variation of 010 superlattice peak due to long time annealing. Data were collected at the BL5S2 beamline at Aichi SR, and the X-ray energy was 20 keV. (e) Diffraction patterns at 110 K for slow-cooled samples, annealed samples at various temperatures, and quenched samples. Annealed samples were held at a specific temperature for one hour. (f) Temperature dependence of magnetic susceptibility of slow-cooled, annealed, and quenched samples during the temperature increase from the lowest temperature to 300 K, respectively. Annealed samples were obtained by holding at a specific temperature for more than 2 hours. The measurements were performed under a magnetic field of 1 T during the temperature increase process from 5 K to 300 K. Inset shows an example of the evolution of magnetic susceptibility with temperature and time.}
\end{figure}

Figures~\ref{fig:fig5}(a-d) show the results of long time annealing, stopping the temperature increase at 150 K, 155 K, 160 K and 165 K when the Ag ordering process started. At each temperature, the peak intensity increases with annealing, indicating that Ag ordering progresses in a time-dependent manner. Regardless of the annealing temperature, after a sufficient time, the intensity becomes constant and Ag ordering reaches thermal equilibrium. The peak intensity at thermal equilibrium varies with annealing temperature, but there is no clear change in the full width at half maximum (FWHM) of the peak with time. This indicates that the correlation length of Ag ordering is determined not by time but by temperature. Although the Ag occupancy at each site is expected to vary with annealing temperature, the broad superlattice peaks make Rietveld analysis inapplicable, and the Ag occupancy at each site cannot be estimated. As shown in Figure~\ref{fig:fig5}(e), the diffraction pattern of the sample that was annealed and then slowly cooled to 110 K is clearly different from that of the sample that was quenched or slowly cooled from room temperature, and the difference in superlattice peak intensity derived from the annealing temperature is retained. This indicates that by changing the annealing temperature, various metastable phases with different Ag site occupancies are realized and maintained at lower temperatures.

Since the ordering of Ag has a significant effect on the charge ordering of V and even dimer formation at low temperatures, annealed samples are expected to exhibit different low-temperature magnetic properties depending on the annealing temperature. In order to clarify how the low-temperature ground state changes with the degree of Ag ordering, magnetic susceptibility measurements were performed on samples annealed at different temperatures for sufficient time. The results are shown in Figure~\ref{fig:fig5}(f). The sample cooled slowly from room temperature shows a rapid decrease in susceptibility due to the spin-singlet transition below 100 K, while the quenched sample shows no decrease in susceptibility below 100 K and a Curie paramagnetic-like increase in susceptibility at the lowest temperature. Calculating the free spin number $N$ from this Curie paramagnetic component, we obtain $N$ = 2.59 $\times$ 10$^{18}$ (per V$^{4+}$), which is about 1/10 of the value expected if all V$^{4+}$ ions generated spin $S$ = 1/2. A similar phenomenon has been reported in previous studies and is argued to be due to the itinerant nature of the system \cite{Ag2/3V2O5}.

Samples annealed after quenching from room temperature showed a slight decrease in magnetic susceptibility below 100 K and a slight increase in Curie paramagnetic susceptibility at the lowest temperature. One example of the temperature/time dependence of the magnetic susceptibility due to annealing is shown in the inset of Figure~\ref{fig:fig5}(f). The degree of decrease in the magnetic susceptibility below 100 K increases with increasing annealing temperature. The free spin values calculated from the Curie paramagnetic components are $N$ = 1.19 $\times$ 10$^{18}$ (per V$^{4+}$ (150 K anneal)), $N$ = 0.94 $\times$ 10$^{18}$ (per V$^{4+}$ (155 K anneal)) and $N$ = 0.85 $\times$ 10$^{18}$ (per V$^{4+}$ (160 K anneal)), showing a clear decreasing trend with increasing annealing temperature. These results indicate that, while no spin singlet state is formed between V$^{4+}$ with spin $S$=1/2 when the valence of V is randomly frozen by quenching, in the sample annealed after quenching, the partial ordering of Ag and the subsequent charge ordering of V and the subsequent gradual change of the free spin of V$^{4+}$ to a spin singlet state.

\subsection{\label{sec:levelG}Discussion}

\begin{figure}
\includegraphics[width=86mm]{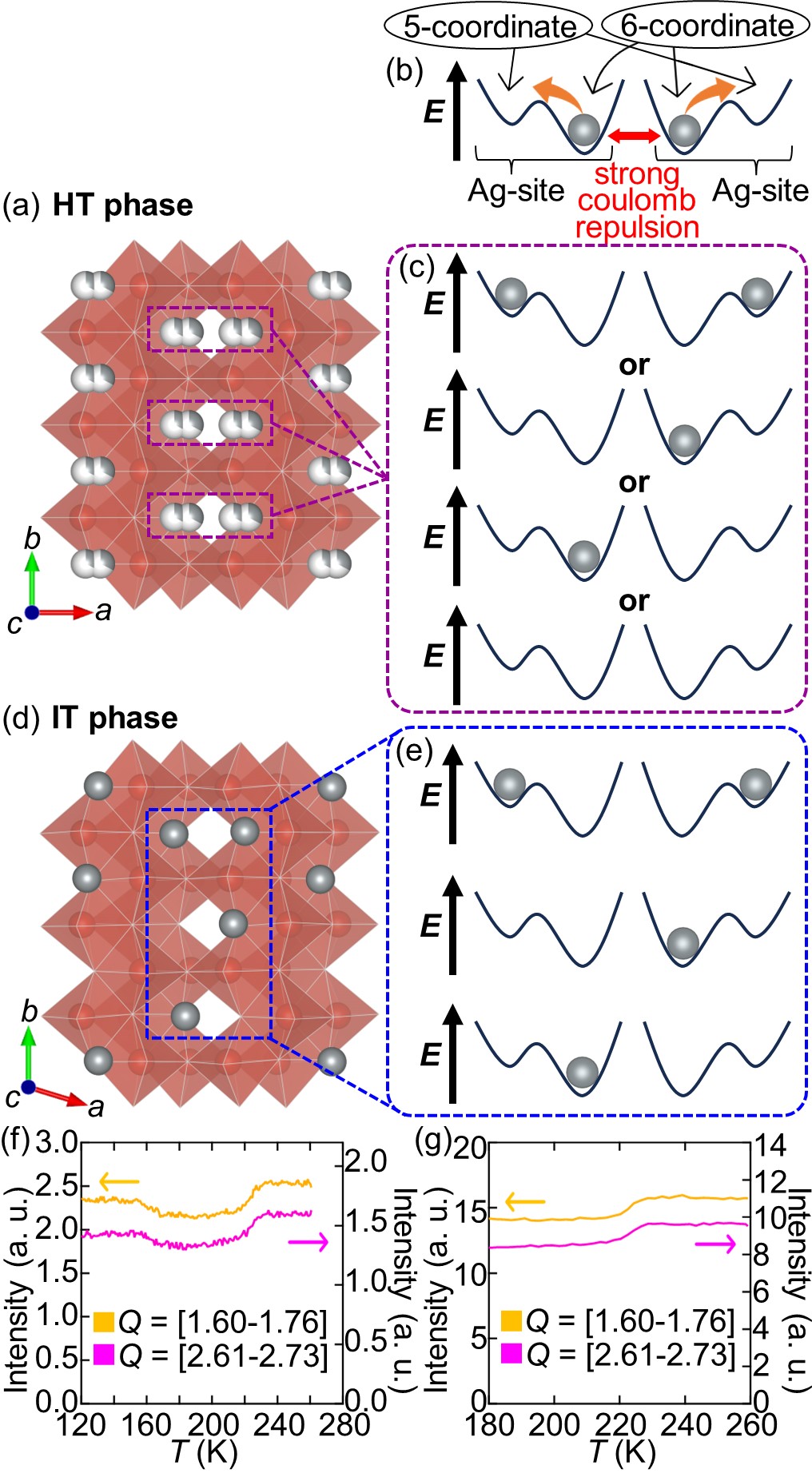}
\caption{\label{fig:fig6} (a) Schematic of the relationship between Ag ion sites with atomic displacements in the high-temperature phase. (b) Relationship between the energies of the 5- and 6-coordination sites, where the 6-coordination site has a shorter distance from the neighboring Ag site and is expected to have strong Coulomb repulsion. (c) Pattern of four Ag ion configurations expected to occur in two adjacent Ag site pairs. (d) Schematic of the relationship between Ag ion sites with lost atomic displacement in the intermediate temperature phase. (e) Patterns of three Ag ion configurations occurring in two adjacent Ag site pairs. The realization of these three patterns is confirmed by Rietveld analysis, and the reason why both 6- and 5-coordinated sites appear can be understood based on the Coulomb repulsion argument. (f) The temperature dependence of the integral of the background data during slow cooling process. (g) The temperature dependence of the integral of the background data during slow heating process of the quenched sample.}
\end{figure}

The above analysis reveals that localized atomic displacement occurs at each Ag site in the high-temperature phase. As shown in Figure~\ref{fig:fig1}(e), these two split Ag ion environments are not identical, but are composed of two sites, one with 5-coordinated oxygen and one with 6-coordinated oxygen. Both of these 5- and 6-coordination sites are too close together for the atoms to enter at the same time. Therefore, as shown in Figure~\ref{fig:fig6}(b), this pair of 5- and 6-coordination sites is referred to as ``Ag-site'' in the following. In the case of 5-coordination and 6-coordination, Ag ions entering the 6-coordination site, where the space created by the oxygen ions is larger, seems to have lower energy in the system. Therefore, as shown in Figure~\ref{fig:fig6}(b), all Ag ions are expected to enter the 6-coordination site and stabilize. However, when both Ag ions enter the 6-coordination site, the distance between the Ag ions should shorten and a strong Coulomb repulsion should occur between the Ag ions. Assuming that this Coulomb repulsion destabilizes the 6-coordination state and induces the 5-coordination state, as shown in the top panel of Figure~\ref{fig:fig6}(c), all the structural analysis results of this system can be explained without contradiction. It is important to note that since the occupancy of the Ag sites in the high-temperature phase is 2/3, there are cases where no Ag is present in one of the two neighboring sites, or no Ag ions in either site. When Ag ion is absent at one of the sites, there is no strong Coulomb repulsion from the neighboring Ag site, as shown in the middle figure of Figure~\ref{fig:fig6}(c), and the Ag ion can enter the energetically stable 6-coordination site. Thus, the effect of 5-coordination induced by Coulomb repulsion explains why Ag ions are present at both the 5-coordination and 6-coordination positions. Furthermore, based on the above discussion, with the 2/3 occupancy of Ag ions in mind, the ratio of Ag entering the 6-coordination site to that entering the 5-coordination site is calculated to be 2:1, which is consistent with the ratio of Ag shown in Figure~\ref{fig:fig1}(e). Similar differences in cation site occupancy as in $\delta$-Ag$_{2/3}$V$_2$O$_5$ have been reported for $\beta'$-Cu$_{x}$V$_2$O$_5$, which belongs to vanadium bronze, and a similar argument may be applicable \cite{CuxV2O5-1, CuxV2O5-2}.

The four patterns shown in Figure~\ref{fig:fig6}(c) occur randomly at the Ag-sites in the high-temperature phase. On the other hand, as shown in Figure~\ref{fig:fig6}(d), as the temperature is lowered, the occupancy of Ag ions becomes more ordered, and at the same time the local displacement at each Ag-site disappears, as shown in the discussion of the temperature factor in Figure~\ref{fig:fig2}(c). The coordinates of each Ag ion in the intermediate temperature phase reveal that each Ag ion is located at the site shown in Figure~\ref{fig:fig6}(e). The reason for the arrangement of Ag ions in the intermediate phase as shown in Figure~\ref{fig:fig6}(e) is easily understood based on the above-mentioned discussion of Coulomb repulsion, and supports the argument that strong Coulomb repulsion between adjacent Ag-sites promotes stabilization of the energetically unstable five-coordination sites. The presence of strong Ag-site disorder in the high temperature phase and Ag-site ordering in the intermediate temperature phase can also be observed indirectly from the color plot in Figure~\ref{fig:fig2}(a). Background data in the region of about $Q$ = 1.6-1.7 and $Q$ = 2.6-2.7 show a clear intensity change around 220 K. The temperature dependence of the integral of the background data in this $Q$ region is summarized in Figure~\ref{fig:fig6}(f). This may indicate that the diffuse streaks originating from the strong Ag site disorder that appear in the high temperature phase disappear in the intermediate temperature phase. A similar trend can be seen in the color plots of Figure~\ref{fig:fig4}(a) and the temperature dependence of the integrated background intensity shown in Figure~\ref{fig:fig6}(g).

As described above, the high-temperature phase has the four Ag ion configurations described in Figure~\ref{fig:fig6}(c) between adjacent Ag sites, which appear randomly. The quench freezes the high-temperature phase state because the thermal energy is too low to jump between different Ag sites. Thus, 160 K, where the superlattice peak begins to develop as the temperature rises, corresponds to the minimum energy required to jump between Ag sites. However, energy splitting within a single Ag-site originating from 5- and 6-coordination complicates the energy required for site-to-site jumps. This is thought to be the reason why the degree of Ag ordering changes gradually with temperature, and even if annealed for a sufficient amount of time at each temperature, it does not order to the same state and different metastable states are realized. In turn, this may be the cause of the different valence separations and partial dimer formation of V, which originate from ordering at the Ag site.

This may suggest that the low structural symmetry of $\delta$-Ag$_{2/3}$V$_2$O$_5$ is the source of its functionality: as shown in Figure~\ref{fig:fig1}(d), there are two locally displaced Ag sites with different coordination environments, at most $\sim$0.6 \AA ~apart, and that these sites are not connected by mirror symmetry and feel different Coulomb repulsion, both of which are difficult to achieve in crystals with a highly symmetric structure. These low symmetry-derived properties and functions may appear relatively universally in vanadium oxides. For example, in $\beta$-Cu$_2$V$_2$O$_7$, the large deformation of the lattice structure with temperature change plays a major role in the realization of negative thermal expansion properties over a wide temperature range. The mechanism of lattice deformation is not clear, but it is expected that the large displacement is caused by the low symmetry of the crystal and few symmetry operations connecting the sites, the oxygen sites sandwiched between the CuO$_4$ tetrahedra, which are the key to lattice deformation, are located at sites with low symmetry and all $x$, $y$ and $z$ degrees of freedom in the fractional coordinates \cite{Cu2V2O7-1, Cu2V2O7-2, Cu2V2O7-3}.

In today's condensed matter physics, many fascinating physical phenomena occur in the vicinity of quantum critical points and other phase transitions, and an important aspect of capturing these physical phenomena is the development of techniques for tuning the electronic state of the system. In many cases, chemical substitution, physical pressure, and electric field effects are used as such techniques, but the results of this study clearly show that it is possible to control the physical properties of the system by freely controlling the order of Ag, the intercalant. The high ionicity and mobility of group 11 elements in the monovalent cationic state, the presence of multiple Ag sites with different potentials due to low crystal symmetry, and the lattice framework of vanadium oxides with a large mobile space are the background that allows such control in $\delta$-Ag$_{2/3}$V$_2$O$_5$. Considering that the role of Ag$^+$ can be replaced by Cu$^+$, Li$^+$, Na$^+$, K$^+$, etc. \cite{Cu2V2O7-2, Cu2V2O7-1}, and that similar low-density crystal structures can be realized in other low-dimensional compounds, the phase control of creating various metastable phases by annealing is a universal technique that can be realized in other material systems. In the future, it is expected that many physical properties and functions will be created through further exploration of physical properties based on this research.

\section{\label{sec:level3}Summary}
In summary, we have performed synchrotron X-ray diffraction and magnetic susceptibility measurements on $\delta$-Ag$_{2/3}$V$_2$O$_5$ for the low temperature state. Synchrotron X-ray structure analysis revealed that slow cooling causes Ag sites to order, resulting in V$^{4+}$/V$^{5+}$ ordering of V and subsequent dimer formation between V$^{4+}$. When the sample is quenched, Ag is frozen in a random position, but subsequent annealing at around 160 K results in the realization of a metastable phase in which Ag is partially ordered. Depending on the degree of Ag ordering, a continuous change in magnetic susceptibility was found to occur. This indicates that Ag ordering can be controlled by annealing and that partial ordering of the glassy Ag sites is an effective means of controlling the charge ordering state and physical properties of $\delta$-Ag$_{2/3}$V$_2$O$_5$.

$\newline$

\begin{acknowledgments}
The work leading to these results has received funding from the Grant in Aid for Scientific Research (Nos.~JP21K18599, JP21J21236, JP23KJ1521, JP23H04104), Research Foundation for the Electrotechnology of Chubu, The Thermal and Electric Energy Technology Inc.. This work was carried out under the Visiting Researcher’s Program of the Institute for Solid State Physics, the University of Tokyo, and the Collaborative Research Projects of Laboratory for Materials and Structures, Institute of Innovative Research, Tokyo Institute of Technology. Research was carried out in part at beamline X17B1 of the National Synchrotron Light Source, Brookhaven National Laboratory, which is supported by the U.S. Department of Energy, Contract No. DE-AC02-76CH00016. Powder XRD experiments were conducted at the BL02B2, BL04B2 and BL19B2 of SPring-8, Hyogo, Japan (Proposals No. 2022B1570, 2022B1130, 2023A1110, 2023A1869), and at the BL5S2 of Aichi Synchrotron Radiation Center, Aichi Science and Technology Foundation, Aichi, Japan (Proposals No. 202205005, 202206142, 202301056 and 202305060).
\end{acknowledgments}

\bibliography{references}

\end{document}